\documentclass[11pt]{iopart}
\usepackage[utf8]{inputenc}
\usepackage{hyperref}
\usepackage{multicol}
\usepackage{xspace}
\expandafter\let\csname equation*\endcsname=\relax
\expandafter\let\csname endequation*\endcsname=\relax
\usepackage[intlimits]{amsmath}
\usepackage{amssymb,amsfonts,amsthm}

\catcode`,\active

\catcode`\,12

\catcode`,\active

\catcode`\,12

\newcommand{\nombrespacs}[1]{\par\noindent
{\small{PACS numbers\/}: #1}}
\newcommand{\class}[1]{\par\noindent
{\small{AMS classification scheme numbers\/}: #1}}

\newcommand{\pd}{\partial}
\newcommand{\ket}[1]{|#1\rangle\xspace}

\newcommand{\bra}[1]{\langle #1|\xspace}

\newcommand{\Braket}[3]{\bra{#1}#2\ket{#3}\xspace}
\newcommand{\braket}[2]{\langle #1|#2\rangle\xspace}

\numberwithin{equation}{section}
\begin{document}
\title[$d$-variate Charlier polynomials as matrix elements of $E(d)$]{The multivariate Charlier polynomials as matrix elements of the Euclidean group representation on oscillator states}
\author{Vincent X. Genest}
\ead{genestvi@crm.umontreal.ca}
\address{Centre de recherches math\'ematiques, Universit\'e de Montr\'eal, Montr\'eal, Qu\'ebec, Canada, H3C 3J7}
\author{Hiroshi Miki}
\ead{hmiki@mail.doshisha.ac.jp}
\address{Department of Electronics, Faculty of Science and Engineering, Doshisha University, Kyotanabe City, Kyoto 610 0394, Japan}
\author{Luc Vinet}
\ead{luc.vinet@umontreal.ca}
\address{Centre de recherches math\'ematiques, Universit\'e de Montr\'eal, Montr\'eal, Qu\'ebec, Canada, H3C 3J7}
\author{Alexei Zhedanov}
\ead{zhedanov@yahoo.com}
\address{Donetsk Institute for Physics and Technology, Donetsk 83114, Ukraine}
\begin{abstract}
A family of multivariate orthogonal polynomials generalizing the standard (univariate) Charlier polynomials is shown to arise in the matrix elements of the unitary representation of the Euclidean group $E(d)$ on oscillator states. These polynomials in $d$ discrete variables are orthogonal on the product of $d$ Poisson distributions. The accent is put on the $d=2$ case and the group theoretical setting is used to obtain the main properties of the polynomials: orthogonality and recurrence relations,  difference equation, raising/lowering relations, generating function, hypergeometric and integral representations and explicit expression in terms of standard Charlier and Krawtchouk polynomials. The approach is seen to extend straightforwardly to an arbitrary number of variables. The contraction of $SO(3)$ to $E(2)$ is used to show that the bivariate Charlier polynomials correspond to a limit of the bivariate Krawtchouk polynomials.

\nombrespacs{02.20.-a, 02.10.Yn, 02.10.De}

\class{33C50, 06B15}
\end{abstract}
\section{Introduction}
In this paper, a new family of multi-variable Charlier polynomials that arise as matrix elements of the unitary reducible Euclidean group representation on oscillator states is introduced. The main properties of these polynomials are obtained naturally from the group theoretical context. The focus is put mainly on the bivariate case, for which the two-variable Charlier polynomials occur in the matrix elements of unitary reducible $E(2)$ representations on the eigenstates of a two-dimensional isotropic harmonic oscillator. The extension to an arbitrary number of variables is straightforward and is given towards the end of the paper.

The standard Charlier polynomials $C_{n}(x;a)$ of degree $n$ in the variable $x$ were introduced in 1905 \cite{Charlier-1905}. These polynomials form one of the most elementary family of orthogonal polynomials (OPs) in the Askey scheme of hypergeometric OPs \cite{Koekoek-2010}. They are orthogonal with respect to the Poisson distribution $w_{x}^{(a)}$ with parameter $a>0$ which is defined by
\begin{equation*}
w_{x}^{(a)}=\frac{a^{x}e^{-a}}{x!},
\end{equation*}
and their discrete orthogonality relation reads
\begin{equation*}
\sum_{x=0}^{\infty}w_{x}^{(a)}C_{n}(x;a)C_{m}(x;a)=a^{-n}n!\,\delta_{nm}.
\end{equation*}
They can be defined through their generating function
\begin{align}
\label{Gen-Charlier}
e^{t}\left(1-\frac{t}{a}\right)^{x}=\sum_{n=0}^{\infty}\frac{C_{n}(x;a)}{n!} t^{n},
\end{align}
and can be expressed in terms of a ${}_2F_{0}$ hypergeometric function (see \cite{Koekoek-2010} for additional properties and references).
The Charlier polynomials appear in various fields including combinatorics \cite{Labelle-1989} as well as statistics and probability \cite{Gessel-2010, Mehta-1988}. In Physics, the importance of these polynomials is mostly due to their appearance in the matrix elements of unitary irreducible representations of the one-dimensional oscillator group \cite{Granovskii-1986, Vilenkin-1991}.

In a recent series of papers \cite{Genest-2014-01, Genest-2013-06, Genest-2013-07-2}, we have presented group theoretical interpretations for two families of multivariate orthogonal polynomials: the multi-variable Krawtchouk and Meixner polynomials. These two families, introduced by Griffiths in \cite{Griffiths-1971-04,Griffiths-1975-06}, were seen to occur in the matrix elements of reducible unitary representations of the rotation and pseudo-rotation groups on oscillator states. This algebraic framework led to a number of new identities for these polynomials and allowed for simple derivations of their known properties.

Here we consider the Euclidean group $E(d)$ which is the group generated by the translations and the rotations in $d$-dimensional Euclidean space.  We shall investigate the matrix elements of the unitary reducible representation of this group on the eigenstates of a $d$-dimensional isotropic harmonic oscillator and show that these are expressed in terms of a new family of multivariate orthogonal polynomials that shall be identified as multivariate extensions of the standard Charlier polynomials. The main properties of these polynomials will be derived in a simple fashion using the group theoretical interpretation.

The paper is organized as follows. In section 2, the unitary representations of the Euclidean group $E(2)$ are defined. In section 3, it is shown that the matrix elements of these representations are given in terms of bivariate polynomials that are orthogonal with respect to the product of two (independent) Poisson distributions. The duality relation satisfied by these polynomials is discussed in section 4. In section 5, a generating function is obtained and the polynomials are identified as multivariate Charlier polynomials. The generating function is used in section 6 to find an explicit expression for these Charlier polynomials in terms of generalized hypergeometric series. The recurrence relations and the difference equations are given in section 7. In section 8, the matrix elements for one-parameter subgroups are considered and used to obtain an explicit expression for the bivariate Charlier polynomials in terms of standard Charlier and Krawtchouk polynomials. In section 9, an integral representation is given. In section 10, it is shown that the bivariate Charlier polynomials can be obtained from the bivariate Krawtchouk polynomials through a limit process. In section 11, the $d$-dimensional case is treated. A conclusion follows.

\section[Representation of E(2) on oscillator states]{Unitary representations of $E(2)$ on oscillator states}
In this section, the reducible unitary representation of the Euclidean group that shall be used throughout the paper is defined. This representation will be specified on the eigenstates of the two-dimensional isotropic harmonic oscillator.
\subsection{The Heisenberg-Weyl algebra}
Let $a_i$, $a_i^{\dagger}$, $i=1,\,2$, be the generators of the Heisenberg-Weyl algebra satisfying the commutation relations
\begin{equation}
\label{Weyl}
[a_i,a_j^{\dagger}]=\delta_{ij},\quad [a_i,a_j]=0,\quad [a_i^{\dagger},a_{j}^{\dagger}]=0.
\end{equation}
This algebra has a standard representation on the basis vectors
\begin{equation*}
\ket{n_1,n_2}\equiv \ket{n_1}\otimes \ket{n_2},\qquad n_1,n_2=0,1,\ldots,
\end{equation*}
defined by the actions of the generators on the factors of the direct product:
\begin{equation}
\label{Actions}
a_i\ket{n_i}=\sqrt{n_i}\,\ket{n_i-1},\quad a_{i}^{\dagger}\ket{n_i}=\sqrt{n_i+1}\,\ket{n_i+1}.
\end{equation}
In view of the commutation relations \eqref{Weyl} and the actions \eqref{Actions}, the basis vectors $\ket{n_1,n_2}$ can equivalently be written as
\begin{equation}
\label{Equiv}
\ket{n_1,n_2}=\frac{(a_1^{\dagger})^{n_1}(a_2^{\dagger})^{n_2}}{\sqrt{n_1!n_2!}}\ket{0,0}.
\end{equation}
In Cartesian coordinates, the algebra \eqref{Weyl} has the following realization:
\begin{equation}
\label{Realization}
a_i=\frac{1}{\sqrt{2}}\left(x_i+\frac{\pd}{\pd x_i}\right),\quad a_i^{\dagger}=\frac{1}{\sqrt{2}}\left(x_i-\frac{\pd}{\pd x_i}\right),\qquad i=1,2.
\end{equation}
\subsection{The two-dimensional isotropic oscillator}
Consider the Hamiltonian $\mathcal{H}$ governing the isotropic harmonic oscillator in the two-dimensional Euclidean space
\begin{equation}
\label{Hamiltonian}
\mathcal{H}=-\frac{1}{2}\left(\frac{\pd^2}{\pd x_1^2}+\frac{\pd^2}{\pd x_2^2}\right)+\frac{1}{2}(x_1^2+x_2^2).
\end{equation}
Using the realization \eqref{Realization}, the Hamiltonian \eqref{Hamiltonian} can be written as
\begin{equation}
\label{Hamiltonian-2}
\mathcal{H}=a_1^{\dagger}a_1+a_{2}^{\dagger}a_2+1.
\end{equation}
It is seen from \eqref{Actions} that the Hamiltonian \eqref{Hamiltonian-2} is diagonal on the basis vectors $\ket{n_1,n_2}$ with energy eigenvalue $E$ given by:
\begin{equation*}
\mathcal{H}\ket{n_1,n_2}=E\ket{n_1,n_2},\qquad E=n_1+n_2+1.
\end{equation*}
The Schr\"odinger equation $\mathcal{H}\Psi=E \Psi$ associated to the Hamiltonian \eqref{Hamiltonian} separates in the Cartesian coordinates  $x_1$, $x_2$. In these coordinates, the wavefunctions take the form
\begin{equation*}
\braket{x_1,x_2}{n_1,n_2}=\Psi_{n_1}(x_1)\Psi_{n_2}(x_2),
\end{equation*}
with
\begin{align}
\label{Exact-Solution}
\braket{x_i}{n_i}=\Psi_{n_i}(x_i)=\sqrt{\frac{1}{2^{n_i}\,\pi^{1/2}\, n_i!}}\,e^{-x_i^2/2}H_{n_i}(x_i),
\end{align}
where $H_{n}(x)$ denotes the Hermite polynomials \cite{Koekoek-2010}. The wavefunctions $\Psi_{n_i}(x_i)$ satisfy the orthogonality relation
\begin{align}
\label{Orthogonality}
\int_{-\infty}^{\infty}\Psi_{n_i}(x_i)\Psi_{n_i'}(x_i)\,\mathrm{d}x_i=\delta_{n_in_i'}.
\end{align}
\subsection[Representation of E(2) on oscillator states]{The representation of $E(2)$ on oscillator states}
The eigenstates of the two-dimensional quantum harmonic oscillator support a reducible representation of the Euclidean group $E(2)$. We introduce the following notation for the basis vectors:
\begin{align*}
\ket{m,n}\equiv \ket{n_1,n_2},
\end{align*}
so that $m$ and $n$ are identified with $n_1$ and $n_2$, respectively. The $E(2)$ group is generated by two translation operators in the $x_1$ and  $x_2$ directions given by
\begin{align}
\label{Translations}
P_1=i(a_1-a_1^{\dagger}),\quad P_2=i(a_2-a_2^{\dagger}),
\end{align}
and by a rotation generator $J$ which has the expression
\begin{align}
\label{Rotation}
J=i(a_{1}a_{2}^{\dagger}-a_1^{\dagger}a_2).
\end{align}
The generators $P_{1}$, $P_{2}$ and $J$ satisfy the commutation relations of the Euclidean Lie algebra $\mathfrak{e}(2)$ which read
\begin{align}
\label{Euclidean-Lie}
[P_1, P_2]=0,\quad [P_2,J]=i P_1,\quad [J, P_1]=i P_2.
\end{align}
Using the formulas \eqref{Actions}, the actions of the Euclidean generators defined by \eqref{Translations} and \eqref{Rotation} on the eigenstates of the two-dimensional oscillator are easily obtained. The assertion that this representation of the Euclidean group is reducible follows from the observation that the Casimir operator $C$ of $\mathfrak{e}(2)$, which can be written as
\begin{align*}
C=P_1^2+P_2^2,
\end{align*}
does not act, as is directly checked, as a multiple of the identity on $\ket{m,n}$.

We use the following notation. Let $T(\theta,\alpha,\beta)$ be a generic element of the Euclidean group $E(2)$ where $\theta$, $\alpha$ and $\beta$ are real parameters; $T(\theta,\alpha,\beta)$ can be written as
\begin{align*}
T(\theta,\alpha,\beta)=
\begin{pmatrix}
\cos \theta & \sin \theta & \alpha/\sqrt{2}
\\
-\sin \theta & \cos \theta & \beta/\sqrt{2}
\\
0 & 0 & 1
\end{pmatrix},
\end{align*}
and represents the Euclidean move
\begin{align*}
(x_1,x_2,1)^{\top}
\rightarrow
T(\theta,\alpha,\beta)
(x_1,x_2,1)^{\top},
\end{align*}
where $z^{\top}$ stands for transposition. The group multiplication law is provided by the standard matrix product. Consider the unitary representation defined by
\begin{align}
\label{Main}
U(T)=e^{i\alpha P_1}e^{i\beta P_2}e^{i\theta J}.
\end{align}
It is readily checked that $U(T)U^{\dagger}(T)=1$. The transformations of the generators $a_{i}$, $a_{i}^{\dagger}$ under the action of $U(T)$ is given by
\begin{align}
\label{Transformations-1}
\begin{aligned}
U(T) a_1 U^{\dagger}(T)=\cos \theta\,a_1-\sin \theta\,a_2-\alpha \cos \theta+\beta \sin \theta,
\\
U(T) a_2 U^{\dagger}(T)=\sin \theta\,a_1+\cos \theta\,a_2-\alpha \sin \theta-\beta \cos \theta,
\end{aligned}
\end{align}
Similar formulas involving $a_1^{\dagger}$ and $a_2^{\dagger}$ are obtained by taking the complex conjugate of \eqref{Transformations-1}. Since one has $X_i=2^{-1/2}(a_i+a_{i}^{\dagger})$, $i=1,2$, the  transformation laws \eqref{Transformations-1} give for the coordinate operator $(X_1,X_2)$
\begin{align}
\begin{pmatrix}
\widetilde{X}_1
\\
\widetilde{X}_2
\end{pmatrix}
=U(T)
\begin{pmatrix}
X_1
\\
X_2
\end{pmatrix}
U^{\dagger}(T)
=
\begin{pmatrix}
\cos \theta & -\sin \theta
\\
\sin \theta & \cos \theta
\end{pmatrix}
\begin{pmatrix}
X_1
\\
X_2
\end{pmatrix}
+
\begin{pmatrix}
A
\\
B
\end{pmatrix},
\end{align}
where
\begin{align}
\label{AB}
A=-\frac{2}{\sqrt{2}}\big(\alpha \cos \theta-\beta \sin \theta\big),\quad B=-\frac{2}{\sqrt{2}}\big(\alpha \sin \theta+ \beta \cos \theta\big).
\end{align}
One thus has
\begin{align*}
U^{\dagger}(T)\ket{x_1,x_2}=\ket{\widetilde{x}_1,\widetilde{x}_2}=\ket{T^{-1}x_1,T^{-1}x_2},
\end{align*}
where $\widetilde{x}_1$, $\widetilde{x}_2$ are given by
\begin{align}
\label{Coordinate}
\begin{pmatrix}
\widetilde{x}_1
\\
\widetilde{x}_2
\end{pmatrix}
=\begin{pmatrix}
\cos \theta & -\sin \theta
\\
\sin \theta & \cos \theta
\end{pmatrix}
\begin{pmatrix}
x_1
\\
x_2
\end{pmatrix}
+
\begin{pmatrix}
A
\\
B
\end{pmatrix},
\end{align}
with $A$, $B$ given by \eqref{AB}. Moreover, one has $U(TT')=U(T)U(T')$ as should be for a group representation. The inverse transformation formulas
\begin{align}
\label{Transformations-2}
\begin{aligned}
U^{\dagger}(T)a_1 U(T)&=\cos \theta\,a_1+\sin \theta\,a_2+\alpha,
\\
U^{\dagger}(T)a_2 U(T)&=-\sin \theta\,a_1+\cos \theta\,a_2 +\beta.
\end{aligned}
\end{align}
and the Glauber formula \cite{Cohen-1973}:
\begin{align}
\label{Glauber}
e^{\gamma\,(a_i^{\dagger}-a_i)}=e^{-\gamma^2/2}e^{\gamma a_i^{\dagger}}e^{-\gamma a_i},\quad i=1,2,
\end{align}
shall also prove useful in what follows.
\section{The representation matrix elements as orthogonal polynomials}
In this section it is shown that the matrix elements of the unitary representation of $E(2)$ defined in section 2 are expressed in terms of bivariate orthogonal polynomials. The matrix elements of $U(T)$ defined by \eqref{Main} can be written as
\begin{align}
\label{Matrix-Elements}
\Braket{i,k}{U(T)}{m,n}=W_{i,k}\, C_{m,n}(i,k),
\end{align}
where $C_{0,0}(i,k)=1$ and where $W_{i,k}$ is defined by
\begin{align}
W_{i,k}=\Braket{i,k}{U(T)}{0,0}.
\end{align}
To ease the notation, the explicit dependence of $U(T)$ on $T$ shall be dropped.
\subsection[Calculation of W]{Calculation of $W_{i,k}$}
The amplitude $W_{i,k}$ can evaluated by a direct computation. Indeed, since one has $e^{i\theta J}\ket{0,0}=\ket{0,0}$,
it follows that
\begin{align*}
W_{i,k}=\Braket{i,k}{e^{i\alpha P_1}e^{i\beta P_2}e^{i\theta J}}{0,0}=\Braket{i,k}{e^{i\alpha P_1}e^{i\beta P_2}}{0,0}.
\end{align*}
Upon using the Glauber formula \eqref{Glauber} to write
\begin{align*}
e^{i\alpha P_1}=e^{\alpha(a_1^{\dagger}-a_1)}=e^{-\alpha^2/2}e^{\alpha a_{1}^{\dagger}}e^{-\alpha a_{1}},
\quad 
e^{i\beta P_2}=e^{\beta(a_2^{\dagger}-a_2)}=e^{-\beta^2/2}e^{\beta a_{2}^{\dagger}}e^{-\beta a_{2}},
\end{align*}
and the actions \eqref{Actions}, one easily finds
\begin{align}
\label{Weight}
W_{i,k}=e^{-(\alpha^2+\beta^2)/2}\frac{\alpha^i \beta^k}{\sqrt{i!k!}}.
\end{align}
\subsection{Raising relations}
It will now be shown that the functions $C_{m,n}(i,k)$ that appear in the matrix elements \eqref{Matrix-Elements} are polynomials of total degree $m+n$ in the discrete variables $i,k$. This will be done by exhibiting raising relations for $C_{m,n}(i,k)$. Consider the matrix element $\Braket{i,k}{U a_{1}^{\dagger}}{m,n}$. One has on the one hand
\begin{align}
\label{R-1}
\Braket{i,k}{U a_{1}^{\dagger}}{m,n}=\sqrt{m+1}\,W_{i,k}\,C_{m+1,n}(i,k),
\end{align}
and on the other hand, using \eqref{Transformations-1}, one has
\begin{multline}
\label{R-2}
\Braket{i,k}{U a_1^{\dagger}}{m,n}=\Braket{i,k}{U a_1^{\dagger} U^{\dagger}U}{m,n}=\cos \theta \sqrt{i}\, W_{i-1,k}C_{m,n}(i-1,k)
\\
-\sin \theta \sqrt{k}\,W_{i,k-1} C_{m,n}(i,k-1)+(\beta \sin \theta-\alpha \cos \theta) W_{i,k} C_{m,n}(i,k).
\end{multline}
Upon comparing \eqref{R-1} and \eqref{R-2}, one obtains using \eqref{Weight}
\begin{align}
\label{R-3}
\begin{aligned}
\sqrt{m+1}\,C_{m+1,n}(i,k)= &\left(\frac{i}{\alpha}\right) \cos \theta\,C_{m,n}(i-1,k)\\
&-\left(\frac{k}{\beta}\right) \sin \theta\,C_{m,n}(i,k-1)+(\beta \sin \theta-\alpha \cos \theta) C_{m,n}(i,k).
\end{aligned}
\end{align}
Considering instead the matrix element $\Braket{i,k}{U a_{2}^{\dagger}}{m,n}$, one similarly finds
\begin{align}
\label{R-4}
\begin{aligned}
\sqrt{n+1}\,C_{m,n+1}(i,k)= &\left(\frac{i}{\alpha}\right) \sin \theta\,C_{m,n}(i-1,k)\\
&+\left(\frac{k}{\beta}\right) \cos \theta\,C_{m,n}(i,k-1)-(\alpha \sin \theta+\beta \cos \theta) C_{m,n}(i,k).
\end{aligned}
\end{align}
By definition one has $C_{-1,n}(i,k)=C_{m,-1}(i,k)=0$ and $C_{0,0}(i,k)=1$. As a consequence, the formulas \eqref{R-3} and \eqref{R-4} can be used to construct $C_{m,n}(i,k)$ from $C_{0,0}(i,k)$ iteratively. One then observes that $C_{m,n}(i,k)$ are polynomials of total degree $m+n$ in the (discrete) variables $i,k$.
\subsection{Orthogonality relation}
The unitarity of the representation \eqref{Main} and the orthonormality of the basis states leads to an orthogonality relation for the polynomials $C_{m,n}(i,k)$. One has
\begin{align*}
\Braket{m',n'}{U^{\dagger}U}{m,n}=\sum_{i,k=0}^{\infty}\Braket{i,k}{U}{m,n}\Braket{m',n'}{U^{\dagger}}{i,k}=\delta_{mm'}\delta_{nn'}.
\end{align*}
Upon using \eqref{Matrix-Elements} and the reality of the matrix elements in the above equation, the following orthogonality relation is obtained:
\begin{align}
\sum_{i,k=0}^{\infty}w_{i,k}\,C_{m,n}(i,k)C_{m',n'}(i,k)=\delta_{mm'}\delta_{nn'},
\end{align}
where $w_{i,k}$ is the product of two independent Poisson distributions with (positive) parameters $\alpha^2$ and $\beta^2$:
\begin{align}
w_{i,k}=W_{i,k}^2=e^{-(\alpha^2+\beta^2)}\frac{\alpha^{2i}\beta^{2k}}{i!k!}.
\end{align}
\subsection{Lowering relations}
Lowering relations for the polynomials $C_{m,n}(i,k)$ can be obtained by considering the matrix elements $\Braket{i,k}{Ua_i}{m,n}$, $i=1,2$ and proceeding as for the raising relations. From the matrix element $\Braket{i,k}{U a_1}{m,n}$, one finds
\begin{align*}
\begin{aligned}
\sqrt{m}\,C_{m-1,n}(i,k)=&\alpha \cos \theta\, C_{m,n}(i+1,k)
\\
&-\beta \sin \theta\, C_{m,n}(i,k+1)+(\beta \sin \theta-\alpha \cos \theta) C_{m,n}(i,k).
\end{aligned}
\end{align*}
From the matrix element $\Braket{i,k}{U a_2}{m,n}$, one obtains
\begin{align*}
\begin{aligned}
\sqrt{n}\,C_{m,n-1}(i,k)=&\alpha \sin \theta\, C_{m,n}(i+1,k)
\\
&+\beta \cos \theta\, C_{m,n}(i,k+1)-(\alpha \sin \theta+\beta \cos \theta) C_{m,n}(i,k).
\end{aligned}
\end{align*}
\section{Duality}
In this section, a duality relation under the exchange of the variables $i,k$ and the degrees $m,n$ in the polynomials $C_{m,n}(i,k)$ is obtained. Consider the matrix elements $\Braket{i,k}{U^{\dagger}}{m,n}$ and write
\begin{align}
\label{D-1}
\Braket{i,k}{U^{\dagger}}{m,n}=\widetilde{W}_{i,k}\widetilde{C}_{m,n}(i,k),
\end{align}
where $\widetilde{C}_{0,0}(i,k)=1$ and $\widetilde{W}_{i,k}=\Braket{i,k}{U^{\dagger}}{0,0}$. To evaluate the amplitude $\widetilde{W}_{i,k}$, one first observes that the identity $\Braket{i,k}{U^{\dagger}a_i}{0,0}=0$ holds for $i=1,2$. Using the inverse transformation formulas \eqref{Transformations-2}, one obtains the following system of difference equation
\begin{align*}
\begin{aligned}
\cos \theta\,\sqrt{i+1}\,\widetilde{W}_{i+1,k}+\sin \theta\,\sqrt{k+1}\,\widetilde{W}_{i,k+1}+\alpha\,\widetilde{W}_{i,k}=0,
\\
-\sin \theta\,\sqrt{i+1}\,\widetilde{W}_{i+1,k}+\cos \theta\,\sqrt{k+1}\,\widetilde{W}_{i,k+1}+\beta\,\widetilde{W}_{i,k}=0.
\end{aligned}
\end{align*}
It is easily seen that the solution of this system is given by
\begin{align*}
\widetilde{W}_{i,k}=C\,\frac{(\beta\sin\theta-\alpha \cos \theta)^{i}(-\alpha \sin \theta-\beta \cos \theta)^{k}}{\sqrt{i!k!}},
\end{align*}
where $C$ is a constant. The value of $C$ can be determined by the normalization condition
\begin{align*}
1=\Braket{0,0}{U^{\dagger}U}{0,0}=\sum_{i,k=0}^{\infty}\Braket{i,k}{U^{\dagger}}{0,0}\Braket{0,0}{U}{i,k}=\sum_{i,k=0}^{\infty}|\widetilde{W}_{i,k}|^2,
\end{align*}
which gives $C^2=e^{-(\alpha \cos \theta-\beta \sin \theta)^2}e^{-(\alpha \sin \theta+\beta \cos \theta)^2}=e^{-(\alpha^2+\beta^2)}$ and thus
\begin{align}
\widetilde{W}_{i,k}=e^{-(\alpha^2+\beta^2)/2}\frac{(\beta\sin\theta-\alpha \cos \theta)^{i}(-\alpha \sin \theta-\beta \cos \theta)^{k}}{\sqrt{i!k!}}.
\end{align}
Note that $\widetilde{W}_{i,k}$ can also be computed directly (see section 5). Since $U^{\dagger}(T)=U(T^{-1})$, the $\widetilde{C}_{m,n}(i,k)$ are the polynomials corresponding to the inverse transformation $T^{-1}$. For a transformation $T\in E(2)$ specified by the parameters $(\theta,\alpha,\beta)$, the inverse $T^{-1}\in E(2)$ is specified by the parameters $(\widetilde{\theta},\widetilde{\alpha},\widetilde{\beta})$ given by
\begin{align}
\label{Para}
\widetilde{\theta}=-\theta,\quad \widetilde{\alpha}=(\beta\sin \theta-\alpha\cos \theta),\quad \widetilde{\beta}=-(\alpha \sin \theta+\beta \cos \theta).
\end{align}
One can also obtain the matrix element \eqref{D-1} by
\begin{align}
\label{D-2}
\Braket{i,k}{U^{\dagger}}{m,n}=\Braket{m,n}{U}{i,k}^{*}=\Braket{m,n}{U}{i,k}=W_{m,n}C_{i,k}(m,n),
\end{align}
where the reality of the matrix elements \eqref{Matrix-Elements} has been used. Upon combining \eqref{D-1} and \eqref{D-2}, one finds that
\begin{align}
\label{D-3}
C_{i,k}(m,n)=\sqrt{\frac{m!n!}{i!k!}}\left(\frac{\widetilde{\alpha}^i\widetilde{\beta}^k}{\alpha^{m}\beta^{n}}\right)\widetilde{C}_{m,n}(i,k),
\end{align}
where $\widetilde{C}_{m,n}(i,k)$ corresponds to the polynomial $C_{m,n}(i,k)$ with parameters $(\widetilde{\theta},\widetilde{\alpha},\widetilde{\beta})$ given by \eqref{Para}. For the two variable polynomials $S_{m,n}(i,k)$ defined by
\begin{align}
\label{Monic}
C_{m,n}(i,k)=\frac{(-1)^{n+m}}{\sqrt{m!n!}}(\alpha \cos\theta-\beta \sin\theta)^{m}(\alpha\sin \theta+\beta \cos\theta)^{n}S_{m,n}(i,k),
\end{align}
the duality relation \eqref{D-3} takes the elegant form
\begin{align*}
S_{i,k}(m,n)=\widetilde{S}_{m,n}(i,k).
\end{align*}
\section{Generating function}
In this section, a generating function for the bivariate orthogonal polynomials $C_{m,n}(i,k)$ is obtained and is seen to correspond to a multivariate extension of that of the standard Charlier polynomials. Consider the generating series
\begin{align}
F(x,y)=\sum_{m,n=0}^{\infty}W_{i,k}C_{m,n}(i,k)\,\frac{x^{m}y^{n}}{\sqrt{m!n!}}=\sum_{m,n=0}^{\infty}\Braket{i,k}{U}{m,n}\,\frac{x^{m}y^{n}}{\sqrt{m!n!}}.
\end{align}
Using the expression \eqref{Equiv} for the basis vectors $\ket{n_1,n_2}$ and the transformation formulas \eqref{Transformations-1}, one has
\begin{align}
\begin{aligned}
&F(x,y)=\sum_{m,n=0}^{\infty}\frac{x^{m}y^{n}}{\sqrt{m!n!}}\Braket{i,k}{U\,\frac{(a_1^{\dagger})^{m}}{\sqrt{m!}}\frac{(a_2^{\dagger})^{n}}{\sqrt{n!}}}{0,0}
=\sum_{m,n=0}^{\infty}\Braket{i,k}{U\frac{(xa_1^{\dagger})^{m}(ya_2^{\dagger})^{n}}{m!n!}}{0,0}
\\
&\,=\Braket{i,k}{U e^{x a_1^{\dagger}}e^{y a_2^{\dagger}}}{0,0}
=\Braket{i,k}{Ue^{x a_1^{\dagger}}U^{\dagger}Ue^{y a_2^{\dagger}}U^{\dagger}U}{0,0}
\\
&\,=\Braket{i,k}{e^{x\,Ua_1^{\dagger}U^{\dagger}}e^{y\,Ua_2^{\dagger}U^{\dagger}}U}{0,0}
\\
&\,=e^{-x(\alpha \cos \theta-\beta \sin \theta)}e^{-y(\alpha \sin \theta+\beta \cos \theta)}\Braket{i,k}{e^{a_1^{\dagger}(x\cos \theta+y\sin \theta)}e^{a_2^{\dagger}(y\cos \theta-x \sin \theta)}U}{0,0}.
\end{aligned}
\end{align}
Since one has $U\ket{0,0}=e^{-(\alpha^2+\beta^2)/2}e^{\alpha a_1^{\dagger}}e^{\beta a_2^{\dagger}}\ket{0,0}$ by the Glauber formula \eqref{Glauber} and by the actions \eqref{Actions}, one finds
\begin{multline*}
F(x,y)=e^{-(\alpha^2+\beta^2)/2}
\\
\times e^{-x(\alpha \cos \theta-\beta \sin \theta)}e^{-y(\alpha \sin \theta+\beta \cos \theta)}
\Braket{i,k}{e^{a_1^{\dagger}\left(\alpha+x\cos\theta+y\sin\theta\right)}e^{a_2^{\dagger}\left(\beta+y\cos \theta-x\sin \theta\right)}}{0,0},
\end{multline*}
which gives
\begin{multline*}
F(x,y)=e^{-(\alpha^2+\beta^2)/2}
\\
\times e^{-x(\alpha \cos \theta-\beta \sin \theta)}e^{-y(\alpha \sin \theta+\beta \cos \theta)}\frac{(\alpha+x \cos\theta+y\sin \theta)^{i}(\beta+y\cos \theta-x\sin \theta)^{k}}{\sqrt{i!k!}}.
\end{multline*}
Recalling the expression \eqref{Weight} for $W_{i,k}$, the following generating function for the polynomials $C_{m,n}(i,k)$ is obtained:
\begin{align}
\label{Gen-1}
\begin{aligned}
e^{-x(\alpha\cos\theta-\beta \sin\theta)}&e^{-y(\alpha\sin \theta+\beta \cos \theta)}\left[1+\frac{x}{\alpha}\cos\theta+\frac{y}{\alpha}\sin\theta\right]^{i}\left[1-\frac{x}{\beta}\sin\theta+\frac{y}{\beta}\cos\theta\right]^{k}
\\
&
=\sum_{m,n=0}^{\infty}C_{m,n}(i,k)\,\frac{x^{m}y^{n}}{\sqrt{m!n!}}.
\end{aligned}
\end{align}
For the polynomials $S_{m,n}(i,k)$ given by \eqref{Monic}, defining
\begin{align*}
z_1=-x(\alpha \cos \theta-\beta \sin \theta)x,\quad z_2=-y(\alpha\sin \theta+\beta \cos \theta),
\end{align*}
yields the generating function
\begin{align}
\label{Gen-Fun}
e^{z_1+z_2}\left[1+u_{11} z_1+u_{12}z_2\right]^{i}\left[1+u_{21}z_1+u_{22}z_2\right]^{k}=\sum_{m,n=0}^{\infty}\frac{S_{m,n}(i,k)}{m!n!}z_1^{m}z_{2}^{n}
\end{align}
where the parameters $u_{ij}$ are of the form
\begin{align}
\label{u-ij}
\begin{aligned}
u_{11}=\frac{-\cos\theta}{\alpha^2\cos \theta-\alpha\beta \sin \theta},\quad u_{12}=\frac{-\sin \theta}{\alpha^2\sin\theta+\alpha\beta \cos\theta},
\\
u_{21}=\frac{-\sin \theta}{\beta^2\sin\theta-\alpha\beta \cos\theta},\quad u_{22}=\frac{-\cos\theta}{\beta^2\cos\theta+\alpha\beta \sin \theta}.
\end{aligned}
\end{align}
The expression \eqref{Gen-Fun} for the generating function of the polynomials $S_{m,n}(i,k)$ lends itself to comparison with that of the Charlier polynomials \eqref{Gen-Charlier}. It is clear from this that $S_{m,n}(i,k)$ can be identified with multivariate Charlier polynomials.
\section{Explicit expression in hypergeometric series}
In this section, an explicit expression for the bivariate Charlier polynomials $S_{m,n}(i,k)$ in terms of a Gelfan'd-Aomoto hypergeometric series is obtained. Consider the generating relation \eqref{Gen-Fun}. Upon denoting by $F(z_1,z_2)$ the left-hand side of \eqref{Gen-Fun} and using the trinomial expansion, one finds
\begin{align}
\label{above}
F(z_1,z_2)=
e^{z_1+z_2}\sum_{\rho,\sigma,\mu,\nu}
\binom{i}{\rho}\binom{i-\rho}{\sigma}\binom{k}{\mu}\binom{k-\mu}{\nu}u_{11}^{\rho}u_{12}^{\sigma}u_{21}^{\mu}u_{22}^{\nu}\;z_1^{\rho+\mu}z_2^{\sigma+\nu},
\end{align}
where the summation runs over all non-negative values of the indices and where binomial coefficients with negative entries are taken to be zero. Upon expanding the exponential in \eqref{above}, gathering the terms in $z_1^{m}z_2^{n}$ and using the identity $\frac{(-1)^{n}m!}{(m-n)!}=(-m)_{n}$ where $(a)_{n}$ stands for the Pochhammer symbol
\begin{align*}
(a)_{n}=a(a+1)\cdots (a+n-1),\quad (a)_0=1,
\end{align*}
one finds that the bivariate Charlier polynomials can be written as
\begin{align}
\label{Hyper-Geo}
S_{m,n}(i,k)=\sum_{\rho,\sigma,\mu,\nu}\frac{(-m)_{\rho+\mu}(-n)_{\nu+\sigma}(-i)_{\rho+\sigma}(-k)_{\mu+\nu}}{\rho!\sigma!\mu!\nu!}\,u_{11}^{\rho}u_{12}^{\sigma}u_{21}^{\mu}u_{22}^{\nu},
\end{align}
where the $u_{ij}$ are given by \eqref{u-ij}. The series appearing in \eqref{Hyper-Geo} is a special case of Gelfan'd-Aomoto hypergeometric series \cite{Aomoto-1994, Gelfand-1986}. The multivariate orthogonal polynomials of Krawtchouk and Meixner type are also known to admit explicit expressions in terms of these multi-variable generalized hypergeometric series (see \cite{Hoare-2008} and \cite{Iliev-2012}).
\section{Recurrence relations and difference equations}
In this section, the group theoretical framework is exploited to obtain the bispectral properties of the bivariate Charlier polynomials $C_{m,n}(i,k)$.
\subsection{Recurrence relations}
A pair of recurrence relations for the bivariate Charlier polynomials can be obtained by examining the matrix elements $\Braket{i,k}{a_{i}^{\dagger}a_i U}{m,n}$ for $i=1,2$. Consider the case $i=1$ first, one has on the one hand
\begin{align}
\label{Recu-1}
\Braket{i,k}{a_{1}^{\dagger}a_{1}U}{m,n}=i\,W_{i,k}\,C_{m,n}(i,k).
\end{align}
On the other hand, using the unitary of $U$, one has
\begin{align}
\label{Recu-2}
\Braket{i,k}{a_1^{\dagger}a_{1}U}{m,n}=\Braket{i,k}{U\;U^{\dagger}a_1^{\dagger}U\;U^{\dagger}a_1U}{m,n}.
\end{align}
Comparing \eqref{Recu-1} with \eqref{Recu-2} using the transformation rules \eqref{Transformations-2}, one obtains the recurrence relation
\small
\begin{multline*}
i\,C_{m,n}(i,k)=\left[m\cos^2\theta+n\sin^2\theta+\alpha^2\right]C_{m,n}(i,k)
\\
+\alpha \sin\theta\left[\sqrt{n+1}C_{m,n+1}(i,k)+\sqrt{n}C_{m,n-1}(i,k)\right]+\sin\theta\cos\theta \sqrt{n(m+1)} C_{m+1,n-1}(i,k)
\\
+\alpha \cos\theta \left[\sqrt{m+1}C_{m+1,n}(i,k)+\sqrt{m}C_{m-1,n}(i,k)\right]+\sin \theta\cos\theta \sqrt{m(n+1)}C_{m-1,n+1}(i,k).
\end{multline*}
\normalsize
In a similar fashion, one finds from the matrix element $\Braket{i,k}{a_{2}^{\dagger}a_2 U}{m,n}$ a second recurrence relation
\small
\begin{multline*}
k\,C_{m,n}(i,k)=\left[m\sin^2\theta+n\cos^2\theta+\beta^2\right]C_{m,n}(i,k)
\\
+\beta \cos\theta\left[\sqrt{n+1}C_{m,n+1}(i,k)+\sqrt{n}C_{m,n-1}(i,k)\right]-\sin\theta\cos\theta \sqrt{n(m+1)} C_{m+1,n-1}(i,k)
\\
-\beta \sin\theta \left[\sqrt{m+1}C_{m+1,n}(i,k)+\sqrt{m}C_{m-1,n}(i,k)\right]-\sin \theta\cos\theta \sqrt{m(n+1)}C_{m-1,n+1}(i,k).
\end{multline*}
\normalsize
\subsection{Difference equations}
A pair of difference equations can be obtained by considering instead the matrix elements $\Braket{i,k}{U a_{i}^{\dagger}a_{i}}{m,n}$ for $i=1,2$. Taking $i=1$, one has
\begin{align*}
\Braket{i,k}{Ua_{1}^{\dagger}a_1}{m,n}=m\,W_{i,k}\,C_{m,n}(i,k).
\end{align*}
Comparing the above relation with $\Braket{i,k}{U a_{1}^{\dagger}a_1}{m,n}=\Braket{i,k}{U a_{1}^{\dagger}U^{\dagger}\;Ua_1U^{\dagger}U}{m,n}$ using the transformation formulas \eqref{Transformations-1}, one obtains
\begin{align*}
m&\,C_{m,n}(i,k)=\left[i\,\cos^2\theta+k\sin^2\theta+\omega^2\right] C_{m,n}(i,k)
\\
&-\omega \cos\theta \left[\frac{i}{\alpha}\,C_{m,n}(i-1,k)+\alpha\,C_{m,n}(i+1,k)\right]-\frac{i\beta}{\alpha}\cos\theta\sin\theta\,C_{m,n}(i-1,k+1)
\\
&+\omega\sin\theta\left[\frac{k}{\beta}\,C_{m,n}(i,k-1)+\beta\, C_{m,n}(i,k+1)\right]-\frac{k\alpha}{\beta}\cos\theta\sin\theta\,C_{m,n}(i+1,k-1),
\end{align*}
where $\omega=\alpha \cos \theta-\beta \sin \theta$. Starting instead from $\Braket{i,k}{U a_{2}^{\dagger}a_2}{m,n}$, one similarly finds
\begin{align*}
n&\,C_{m,n}(i,k)=\left[i\,\sin^2\theta+k\cos^2\theta+\zeta^2\right] C_{m,n}(i,k)
\\
&-\zeta \sin\theta \left[\frac{i}{\alpha}\,C_{m,n}(i-1,k)+\alpha\,C_{m,n}(i+1,k)\right]+\frac{i\beta}{\alpha}\cos\theta\sin\theta\,C_{m,n}(i-1,k+1)
\\
&-\zeta\cos\theta\left[\frac{k}{\beta}\,C_{m,n}(i,k-1)+\beta\, C_{m,n}(i,k+1)\right]+\frac{k\alpha}{\beta}\cos\theta\sin\theta\,C_{m,n}(i+1,k-1),
\end{align*}
with $\zeta=\alpha \sin\theta+\beta \cos \theta$.
\section{Explicit expression in standard Charlier and Krawtchouk polynomials}
In this section, an explicit expression for the bivariate Charlier polynomials $C_{m,n}(i,k)$ involving the standard (univariate) Charlier and Krawtchouk polynomials. To this end, one first notes that the matrix elements $\Braket{i,k}{U}{m,n}$ can be decomposed as follows:
\begin{align}
\label{Decompo}
\Braket{i,k}{U}{m,n}=\sum_{r,s,u,v=0}^{\infty}\Braket{i,k}{e^{iP_1}}{r,s}\Braket{r,s}{e^{i\beta P_2}}{u,v}\Braket{u,v}{e^{i\theta J}}{m,n}.
\end{align}
We shall now examine individually the matrix elements appearing in the right hand side of the above equation.

Consider the matrix element $\Braket{i,k}{e^{i\alpha P_1}}{r,s}$ which corresponds to a translation in the $x_1$ direction. Since $P_1$ acts only on the first quantum number, one has
\begin{align*}
\Braket{i,k}{e^{i\alpha P_1}}{r,s}=\delta_{ks}\Braket{i,k}{e^{\alpha P_1}}{r,k}.
\end{align*}
The expression for the above matrix element is well known (see for example \cite{VZ-2011-08} or \cite{Vilenkin-1991}) and can be obtained directly by taking $\beta=0$ in the formula \eqref{Weight} for the amplitude $W_{i,k}$ and by taking $\theta=\beta=y=0$ in the generating function \eqref{Gen-1}. Comparing with \eqref{Gen-Charlier}, one finds that
\begin{align}
\label{Exp-1}
\Braket{i,k}{e^{i\alpha P_1}}{r,s}=\delta_{ks}\frac{(-1)^{r}\alpha^{r+i}}{\sqrt{i!r!}}e^{-\alpha^2/2}C_{r}(i;\alpha^2),
\end{align}
where $C_{n}(x;a)$ are the standard Charlier polynomials \cite{Koekoek-2010}. Similarly, one has
\begin{align}
\label{Exp-2}
\Braket{r,s}{e^{i\beta P_2}}{u,v}=\delta_{ru}\frac{(-1)^{v}\beta^{v+s}}{\sqrt{s!v!}}e^{-\beta^2/2}C_{v}(s;\beta^2).
\end{align}
The matrix element $\Braket{u,v}{e^{i\theta J}}{m,n}$ can be evaluated straightforwardly using the methods of \cite{Genest-2014-01, Genest-2013-06}. The result reads
\begin{align}
\label{Exp-3}
\Braket{u,v}{e^{i\theta J}}{m,n}=\delta_{u+v,m+n}(-1)^{v}\binom{N}{v}^{1/2}\binom{N}{n}^{1/2}\cos^{N}\theta\tan^{v+n}\theta\,K_{n}(v;\sin^2\theta;N)
\end{align}
where $u+v=m+n=N$ and where $K_{n}(x;p;N)$ stands for the standard Krawtchouk polynomials \cite{Koekoek-2010}. Upon using the matrix elements \eqref{Exp-1}, \eqref{Exp-2} and \eqref{Exp-3} in the decomposition formula \eqref{Decompo} and using the formula \eqref{Weight}, one finds that the bivariate Charlier polynomials $C_{m,n}(i,k)$ are given by
\begin{multline}
C_{m,n}(i,k)=\frac{(-1)^{n+m}\alpha^{m+n}\cos^{m}\theta\sin^{n}\theta }{\sqrt{m!n!}}
\\
\sum_{v=0}^{m+n}\binom{m+n}{v}\left(\frac{-\beta \sin \theta}{\alpha \cos \theta}\right)^{v}C_{v}(k;\beta^2)C_{m+n-v}(i;\alpha^2)K_{n}(v;\sin^2\theta,n+m),
\end{multline}
where $C_{n}(x;a)$ and $K_{n}(x;p;N)$ are the standard Charlier and Krawtchouk polynomials.
\section{Integral representation}
An integral representation for the bivariate Charlier polynomials can be obtained by considering the matrix element $\Braket{x_1,x_2}{U}{m,n}$ from two different points of view. First, consider the action of $U$ on $\ket{m,n}$. Using the definition \eqref{Matrix-Elements} of the matrix elements, one can write
\begin{align*}
\Braket{x_1,x_2}{U}{m,n}=\sum_{i,k=0}^{\infty}W_{i,k}C_{m,n}(i,k)\,\Psi_{i}(x_1)\Psi_{k}(x_2),
\end{align*}
where $\Psi_{n_i}(x_i)$, $i=1,2$, is given by \eqref{Exact-Solution}. Second, take the action of $U$ on the bra $\bra{x_1,x_2}$. In view of \eqref{Coordinate}, one may write
\begin{align*}
\Braket{x_1,x_2}{U}{m,n}=\braket{\widetilde{x}_1,\widetilde{x}_2}{m,n}=\Psi_{m}(\widetilde{x}_1)\Psi_{n}(\widetilde{x}_2),
\end{align*}
where $(\widetilde{x}_1,\widetilde{x}_2)$ is given by \eqref{Coordinate}. Combining the two previous relations, there comes
\begin{align*}
\Psi_{m}(\widetilde{x}_1)\Psi_{n}(\widetilde{x}_2)=\sum_{i,k=0}^{\infty}W_{i,k}\,C_{m,n}(i,k)\,\Psi_{i}(x_1)\Psi_{k}(x_2).
\end{align*}
Upon multiplying both sides of the above equation by $\Psi_{i'}(x_1)\Psi_{k'}(x_2)$, integrating over the whole Euclidean plane and using the orthogonality relation \eqref{Orthogonality} for the wavefunctions, one finds that
\begin{align*}
C_{m,n}(i,k)=\frac{1}{W_{i,k}}\int_{-\infty}^{\infty}\int_{-\infty}^{\infty}\Psi_{i}(x_1)\Psi_{k}(x_2)\Psi_{m}(\widetilde{x}_1)\Psi_{n}(\widetilde{x_2})\,\mathrm{d}x_1\mathrm{d}x_2,
\end{align*}
with $(\widetilde{x}_1,\widetilde{x}_2)$ given by \eqref{Coordinate}. In view of \eqref{Exact-Solution}, this gives a formula for the bivariate Charlier polynomials $C_{m,n}(i,k)$ in terms of a double integral of a product of four Hermite polynomials.
\section{Charlier polynomials as limits of Krawtchouk polynomials}
In this section, it is shown that the bivariate Charlier polynomials $C_{m,n}(i,k)$ can be obtained from the bivariate Krawtchouk polynomials by a limit process. We begin by providing some background information on the bivariate Krawtchouk polynomials.
\subsection{Bivariate Krawtchouk polynomials}
The bivariate Krawtchouk polynomials $P_{m,n}(i,k;N)$ of two discrete variables $i$, $k$ arise as matrix elements of the rotation group $SO(3)$ on the energy $E=N+3/2$ eigenspace of a three-dimensional isotropic harmonic oscillator \cite{Genest-2013-06}. In addition to the non-negative integer $N$, the bivariate Krawtchouk polynomials have for parameters the entries of a rotation matrix $R\in SO(3)$. Hence for each $N$ and $R\in SO(3)$, one has a finite set of bivariate Krawtchouk polynomials. The polynomials $P_{m,n}(i,k;N)$ can be defined through the following generating function:
\begin{align}
\label{Gen-Fun-Krawtchouk}
\begin{aligned}
G(u,v)=\left(1+\frac{R_{11}}{R_{13}}u+\frac{R_{12}}{R_{13}}v\right)^{i}&\left(1+\frac{R_{21}}{R_{23}}u+\frac{R_{22}}{R_{23}}v\right)^{k}\left(1+\frac{R_{31}}{R_{33}}u+\frac{R_{32}}{R_{33}}v\right)^{N-i-k}
\\
&=\sum_{\substack{m,n=0\\ m+n\leqslant N}}^{N}\binom{N}{m,n}^{1/2}\,P_{m,n}(i,k;N)\,u^{m}v^{n},
\end{aligned}
\end{align}
where $\binom{N}{m,n}$ stands for the trinomial coefficients
\begin{align*}
\binom{N}{m,n}=\frac{N!}{m!n!(N-m-n)!}.
\end{align*}
They satisfy the orthogonality relation
\begin{align*}
\sum_{\substack{i,k=0\\i+k\leqslant N}}^{N}w_{i,k;N} P_{m,n}(i,k;N)P_{m',n'}(i,k;N)=\delta_{mm'}\delta_{nn'},
\end{align*}
with respect to the discrete weight
\begin{align*}
w_{i,k;N}=\binom{N}{i,k}R_{13}^{2i}R_{23}^{2k}R_{33}^{2(N-i-k)}.
\end{align*}
The normalization condition $\sum_{i+k\leqslant N}^{N}w_{i,k;N}=1$ is ensured by the fact that $R$ is an orthogonal matrix, i.e. $RR^{T}=1$.
\subsection[Limit of bivariate Krawtchouk polynomials]{The $N\rightarrow \infty$ limit of the bivariate Krawtchouk polynomials}
It is well known that the $E(2)$ group can be obtained from the $SO(3)$ group by a contraction \cite{Wigner-1953}. Consider the Lie algebra $\mathfrak{so}(3)$ defined by the commutation relations
\begin{align*}
[J_1,J_2]=i J_3,\quad [J_2,J_3]=i J_1,\quad [J_3,J_1]=iJ_2.
\end{align*}
Upon redefining $J_1=\epsilon P_1$, $J_2=\epsilon P_2$ and $J_3=J$, it is easily seen that in the limit as $\epsilon\rightarrow 0$, the $\mathfrak{so}(3)$ commutation relations contract to those of the Euclidean Lie algebra $\mathfrak{e}(2)$ given in \eqref{Euclidean-Lie}. In view of the connection between $SO(3)$ and bivariate Krawtchouk polynomials, this relation can be used to obtain the bivariate Charlier polynomials $C_{m,n}(i,k)$ defined here as limits of the bivariate Krawtchouk polynomials.

Let $R$ be a general $SO(3)$ rotation matrix. One can take the parametrization
\begin{align}
\label{Para-2}
R=r_{x_2}(\delta)r_{x_1}(\gamma)r_{x_3}(\theta),
\end{align}
where $r_{x_i}$, $i=1,2,3$, are the rotation matrices around the $x_1$, $x_2$ and $x_3$ axes:
\small
\begin{gather*}
r_{x_2}(\delta)=
\begin{pmatrix}
\cos \delta & 0 & \sin \delta
\\
0 & 1 & 0
\\
-\sin \delta & 0 & \cos \delta
\end{pmatrix},
\quad 
r_{x_1}(\gamma)=
\begin{pmatrix}
1 & 0 & 0
\\
0 & \cos \gamma & \sin \gamma
\\
0 & -\sin \gamma & \cos \gamma
\end{pmatrix},
\\
 r_{x_3}(\theta)=
\begin{pmatrix}
\cos \theta & \sin \theta & 0
\\
-\sin \theta & \cos \theta & 0
\\
0 & 0 &1
\end{pmatrix}.
\end{gather*}
\normalsize
Upon taking the parametrization
\begin{align}
\label{Para-3}
\delta\rightarrow \frac{\alpha}{\sqrt{N}},\quad \gamma\rightarrow \frac{\beta}{\sqrt{N}},
\end{align}
in the generating function \eqref{Gen-Fun-Krawtchouk} of the bivariate Krawtchouk polynomials, a direct computation shows that 
\begin{multline}
\label{First}
\lim_{N\rightarrow \infty} G\left(\frac{x}{\sqrt{N}},\frac{y}{\sqrt{N}}\right)=
\\
e^{-x(\alpha \cos \theta-\beta \sin \theta)}e^{-y(\alpha \sin \theta-\beta \cos \theta)}\left(1+\frac{x}{\alpha}\cos \theta+\frac{y}{\alpha}\sin \theta\right)^{i}\left(1-\frac{x}{\beta}\sin \theta+\frac{y}{\beta}\cos \theta\right)^{k},
\end{multline}
and also that
\begin{align}
\label{Second}
\lim_{N\rightarrow \infty} G\left(\frac{x}{\sqrt{N}},\frac{y}{\sqrt{N}}\right)=\sum_{m,n=0}^{\infty}\frac{x^{m}y^{n}}{\sqrt{m!n!}}\lim_{N\rightarrow \infty}P_{m,n}(i,k;N).
\end{align}
Combining \eqref{First} and \eqref{Second}, it is directly seen that the resulting generating function coincides with that of the bivariate Charlier polynomials given by \eqref{Gen-1}. Consequently, under the parametrizations \eqref{Para-2} and \eqref{Para-3}, the Charlier polynomials $C_{m,n}(i,k)$ can be obtained by a $N\rightarrow \infty$ limit of the bivariate Krawtchouk polynomials $P_{m,n}(i,k;N)$:
\begin{align}
\lim_{N\rightarrow \infty} P_{m,n}(i,k;N)=C_{m,n}(i,k).
\end{align}
This limiting procedure can be applied to the raising/lowering relations, difference equations, recurrence relations and explicit expression derived in \cite{Genest-2013-06} for the bivariate Krawtchouk polynomials and it is verified that these yield the corresponding relations obtained here for the bivariate Charlier polynomials.
\section{Multidimensional case}
In this section, it is shown how the results obtained so far can be generalized by considering the space of state vectors of a $d$-dimensional isotropic harmonic oscillator to obtain an algebraic description of the multivariate Charlier polynomials in $d$ variables. Consider the Hamiltonian of an isotropic $d$-dimensional harmonic oscillator
\begin{align*}
\mathcal{H}=\sum_{k=1}^{d}a_{k}^{\dagger}a_{k}+d/2,
\end{align*}
where the creation/annihilation operators  $a_{k}^{\dagger}$ , $a_{k}$ satisfy the Weyl algebra commutation relations \eqref{Weyl}. An eigenbasis for $\mathcal{H}$ is provided by the state vectors $\ket{n_1,n_2,\ldots,n_{d}}$ where $n_{k}$, $k=1,\ldots,d$, are non-negative integers and the action of the creation/annihilation operators is given by \eqref{Actions}. The states $\ket{n_1,n_2,\ldots,n_d}$ provide a reducible representation of the Euclidean group $E(d)$. The elements of the Euclidean group $E(d)$ are specified by a $d\times d$ orthogonal matrix $R$ and a real vector $(\alpha_1,\alpha_2,\ldots,\alpha_{d})$ with $d+1$ components. These elements denoted by $T(R,\alpha)$ hence depend on $\frac{d(d+1)}{2}$ independent parameters and they can be represented by the $(d+1)\times (d+1)$ matrix
\begin{align*}
T(R,\alpha)=
\begin{pmatrix}
R & 
\begin{pmatrix}
\alpha_1/\sqrt{2}
\\ 
\vdots
\\
\alpha_d/\sqrt{2}
\end{pmatrix}
\\
\mathbf{0}& 1
\end{pmatrix}.
\end{align*}
The group law is provided by matrix multiplication. Consider now the unitary representation $U(T)$ defined by
\begin{align*}
U(T)=\prod_{k=1}^{d}e^{\alpha_k(a_k^{\dagger}-a_{k})}\,e^{\sum_{j,k=1} B_{jk}a_{j}^{\dagger}a_{k}},
\end{align*}
where the rotation matrix $R$ is related to the antisymmetric matrix $B$ by $e^{B}=R$. The transformations of the generators $a_{k}$, $a_{k}^{\dagger}$ under the action $U(T)$ are
\begin{align}
U(T)a_{k}U^{\dagger}(T)=\sum_{j=1}^{d}R_{jk}\left(a_{j}^{\dagger}+\alpha_j\right),
\end{align}
and similarly for $a_{k}^{\dagger}$. In the same spirit as in section 3, one can write the matrix elements of this reducible $E(d)$ representation on the eigenstates of the $d$-dimensional isotropic oscillator as follows
\begin{align*}
\Braket{i_1,i_2,\ldots,i_{d}}{U(T)}{n_1,n_2,\ldots,n_{d}}=W_{i_1,\ldots,i_{d}}C_{n_1,\ldots,n_{d}}(i_1,\ldots,i_{d}),
\end{align*}
where
\begin{align*}
W_{i_1,\ldots,i_{d}}=\Braket{i_1,\ldots,i_{d}}{U(T)}{0,\ldots,0}.
\end{align*}
Since $U(T)\ket{0,\ldots,0}=\prod_{k=1}^{d}e^{\alpha_k(a_k^{\dagger}-a_{k})}\ket{0,\ldots,0}$, the amplitude $W_{i_1,\ldots,i_{d}}$ is directly evaluated to
\begin{align*}
W_{i_1,\ldots,i_{d}}=e^{-\sum_{k=1}^{d}\alpha_k^2/2}\prod_{k=1}^{d}\frac{\alpha_k^{i_k}}{\sqrt{i_k !}}.
\end{align*}
It is easily verified by deriving the raising relations as in section 3, that the $C_{n_1,\ldots,n_{d}}(i_1,\ldots,i_{d})$ are polynomials in the discrete variables $i_1,\ldots,i_{d}$ of total degree $n_1+n_2+\cdots+n_{d}$. These polynomials are orthogonal with respect to the product of $d$ independent Poisson distributions
\begin{align*}
\sum_{i_1,\ldots i_{d}=0}^{\infty}W_{i_1,\ldots,i_{d}}^2 C_{n_1,\ldots,n_{d}}(i_1,\ldots,i_{d})C_{m_1,\ldots,m_{d}}(i_1,\ldots,i_{d})=\delta_{n_1m_1}\delta_{n_2m_2}\cdots \delta_{n_{d}m_{d}}.
\end{align*}
The generating function can be obtained following the method of section 5 and one finds 
\begin{align*}
e^{-\sum_{i,j}R_{ij}\alpha_ix_j}\prod_{k=1}^{d}\left(1+\sum_{\ell=1}^{d}R_{k\ell}\frac{x_\ell}{\alpha_k}\right)^{i_k}=\sum_{n_1,\ldots,n_{d}=0}^{\infty}C_{n_1,\ldots,n_{d}}(i_1,\ldots,i_{k})\,\frac{x_1^{n_1}x_{2}^{n_2}\cdots x_{k}^{n_k}}{\sqrt{n_1!n_2!\cdots n_{k}!}}
\end{align*}
Deriving the properties of the $d$-variable polynomials $C_{n_1,\ldots,n_{d}}(i_1,\ldots,i_{d})$ can be done exactly as for $d=2$.
\section{Conclusion}
In this paper we have considered the matrix elements of the unitary representation of the Euclidean group $E(2)$ on the states of the two-dimensional isotropic harmonic oscillator and showed that these matrix elements can be expressed in terms of new bivariate orthogonal polynomials that generalize the standard Charlier polynomials. Using the group theoretical setting, the main properties of the polynomials were derived. Furthermore, it was shown that the approach easily extends to $d$ dimensions giving the $d$-variate Charlier polynomials as matrix elements of unitary representations of the Euclidean group $E(d)$ on oscillator states. Let us now offer some comments.

As a first remark, we note that the approach proposed here could be modified straightforwardly to obtain a different family of multivariate Charlier polynomials associated to the pseudo-Euclidean group $E(d-1,1)$. In the bivariate case, the $\mathfrak{e}(1,1)$ generators can be realized with the creation/annihilation operators in the following way:
\begin{align*}
\widetilde{P}_1=i(a_1-a_1^{\dagger}),\quad \widetilde{P}_2=i(a_2-a_2^{\dagger}),\quad K=i(a_1^{\dagger}a_2^{\dagger}-a_{1}a_{2}).
\end{align*}
Using this realization, the unitary representation of the group $E(1,1)$ on the states $\ket{n_1,n_2}$ can be constructed as in section 2 and the matrix elements can be expressed in terms of multivariate orthogonal polynomials.

As a second remark, it is worth mentioning that the results presented here can be combined with those of \cite{Genest-2014-01} to construct unitary representations of the Poincar\'e group on oscillator states whose matrix elements are given in terms of both multivariate Charlier and Meixner polynomials. In two dimensions the Poincar\'e group generators are realized as follows with the operators of three harmonic oscillators (see \cite{Genest-2014-01}):
\begin{align*}
& \text{Space translations:}\quad P_1=\frac{i}{\sqrt{2}}(a_{1}^{\dagger}-a_{1}),\quad P_2=\frac{i}{\sqrt{2}}(a_2^{\dagger}-a_{2}),
\\
&\text{Time translation:}\quad P_0=\frac{i}{\sqrt{2}}(a_{3}-a_{3}^{\dagger}),
\\
&\text{Lorentz Boosts:}\quad K_1=i(a_{1}^{\dagger}a_{3}^{\dagger}-a_{1}a_{3}),\quad K_2=i(a_2^{\dagger}a_3^{\dagger}-a_{2}a_3),
\\
&\text{Rotation:}\quad J=i(a_1a_2^{\dagger}-a_1^{\dagger}a_2).
\end{align*}
For this case, it would be of interest to proceed as in \cite{Genest-2013-07-2} and decompose the unitary representations in its irreducible components.

As a last remark, it is observed that our approach offers a path to defining $q$-extensions of the multi-variable Charlier polynomials considered here. Indeed, one could consider the realization of the quantum group $E_{q}(2)$ with two $q$-oscillators and construct the matrix elements of $q$-exponentials in the $E_{q}(2)$ generators. A comparison with the multivariate $q$-Charlier polynomials defined in \cite{Gasper-2007} would be of interest. We hope to report on this in the future. 
\section*{Acknowledgments}
V.X.G. holds an Alexander-Graham-Bell fellowship from the Natural Sciences and Engineering Research Council of Canada (NSERC). The research of L.V. is supported in part by NSERC.
\phantom{\cite{Rosengren-1998,Iliev-2007,Xu-2004}}
\section*{References}
\footnotesize

\end{document}